# Glass Transition Behavior of Polymer Films of Nanoscopic Dimensions

Arlette R.C. Baljon[*], Maarten H.M. Van Weert, Regina Barber DeGraaff, and Rajesh Khare[**]

*Department of Physics, San Diego State University, San Diego, CA 92182, USA*

(Dated: June 15, 2004)

Glass transition behavior of nanoscopically thin polymer films is investigated by means of molecular dynamics simulations. We study thin polymer films composed of bead-spring model chains and supported on an idealized FCC lattice substrate surface. The impact on the glass transition temperature of the strength of polymer-surface interaction, and of chain grafting to the surface is investigated. Three different methods - volumetric, energetic, and dynamic – are used to determine the glass transition temperature of the films. Based on these, we are able to distinguish two different transition temperatures. When the temperature is lowered, a first transition occurs when the beads order locally. This transition is characterized by an anomaly in the heat capacity. Upon decreasing the temperature further, the point is reached at which internal relaxation times diverge, as calculated, using for instance mode coupling theory. In qualitative agreement with the experiments, the former temperature depends on the characteristics of the polymer-surface interaction. By contrast, the latter temperature, is independent of these

## I. INTRODUCTION

The properties of polymers start deviating from their bulk values when one of the dimensions of the system approaches nanometer length scales. Thin polymeric films with thickness less then 100 nm play an important role in the microelectronics industry, where they are used as masks in lithographic processes. It is critical that in such operations they retain their patterns. Therefore, they should be so designed as to exhibit a high glass transition temperature at processing conditions. Hence, glass transition behavior of thin polymer films in recent years has been the subject of many experimental studies[1,2,3,,4,5,6,7,8].

Two approaches have been discussed in the literature for raising the glass transition temperature of thin polymer films. The first consists of tuning the interfacial energy between the polymer and the surface supporting the film. In an early study, Van Zanten et al.[3] showed that 100 nm thin films of poly-(2)-vinylpyridine, supported on a silicon oxide substrate, exhibit an increase in $T_g$ by 20-50 deg C compared to its bulk value.

More recently, Fryer et al.[4] presented a systematic experimental study of the effect of interfacial energy on the glass transition behavior of thin polymeric films. They found that for a given film thickness, the difference between the glass transition temperatures of the thin film and the bulk polymer scaled linearly with the interfacial energy. This work also demonstrated that for lithographically relevant film thicknesses, the effect of the interfacial energy would not suffice to cause the desired increase in the glass transition temperature of the film.

A second approach for raising the glass transition temperature of thin polymeric films, has been to graft (attach) some of the chains in the film to the surface.[5,6,7,8] Using optical waveguide spectroscopy, Prucker et al. compared the $T_g$ of grafted films of PMMA on a silicon oxide surface with that for films supported on a hydrophobized silicon oxide surface. They found that chain grafting had a negligible effect on $T_g$. Tsui et al.[6] used X-ray reflectivity to study the glass transition behavior of thin films of polystyrene supported on a silicon oxide surface. Films consisted of polystyrene chains spin-coated onto approximately 3 nm thick brush layer of chains end-grafted on the supporting surface. They found that, for 33 nm thick films, the $T_g$ of the films containing a brush layer was only slightly (less than 3 deg C) higher than the $T_g$ of the films that did not contain any grafted chains. Effect of chain grafting on the glass transition behavior of polystyrene films was also the focus of a study by Tate et al.[7]. These authors studied two types of films: In one type of film, some polystyrene chains in the film (that contained terminal hydroxyl groups) were end-grafted to the supporting silicon oxide surface. The other type of film contained poly (4-hydroxystyrene) chains, some of which were side-grafted to the supporting surface. It was found that for 50 nm thick films of the first type, $T_g$ exceeds the bulk $T_g$ by about 15-20 deg C. For 100 nm thick films of the second type, the $T_g$ was raised above the bulk value by as much as 55 deg C.

Besides experimental work, molecular dynamics simulation studies of glass transition behavior of thin polymeric films have been reported in the literature[9,10]. Torres et al.[9] used hard-sphere molecular dynamics simulations to investigate the glass transition behavior of ultrathin films of short polymeric chains. They found that freestanding films exhibit a reduction in $T_g$ for small film thicknesses. For supported films of thickness less than 30 $\sigma$ (where $\sigma$ is the width of the square well potential used for describing non-bonded interactions between various chain sites in their model), they found that the systems with weakly attractive polymer-substrate interactions showed a decrease in $T_g$, whereas the systems with strongly attractive polymer-substrate interactions showed an increase in $T_g$ compared to the bulk value. Varnik et al.[10] also have used molecular dynamics simulations to study the glass transition behavior of thin films of a non-entangled polymer melt confined between smooth and repulsive walls. They found that the glass transition temperature decreases with film thickness, in agreement with experimental results

In this work, we use a molecular model similar to that of Varnik et al.[10] to investigate the effects of chain grafting and polymer-substrate interactions on the glass transition temperature of thin polymer films. Our work is inspired by the aforementioned experiments. The main objective of the work, however, is not to make a quantitative comparison with these experiments; rather it is to obtain an understanding of the molecular level mechanisms that are responsible for the elevation of the glass transition temperature caused by chain grafting. To our knowledge, none of the existing theories for the glass transition, such as those based on the concept of free volume, can explain such a phenomenon. In this study, we compare results for the glass transition temperature ($T_g$) obtained using three different techniques: measurement of changes in the film thickness (volumetric), calculations of the specific heat (energetic), and measurements of relaxation times from diffusion data (dynamic). In each case, we investigate the dependence of $T_g$ on polymer-substrate interaction and grafting.

The rest of the paper is arranged as follows. First, we present the molecular model and the simulation details employed in the work. Results for the glass transition temperature of the thin polymer films obtained using different approaches are discussed next. The paper ends with a discussion of our results and conclusions.

## Molecular Model and Simulation Details

In this work, we study thin polymer films supported on a substrate. The other surface of the films is free. A bead spring model[11] is used for the polymer chains. The system contains $M = 40$ linear polymer chains, each of which consists of $N = 100$ monomers. This chain length is believed to be about twice the entanglement length at the density simulated in this work[12]. The particles that form the substrate are fixed to their positions in an FCC lattice.

Non-bonded interactions between polymer chain beads are modeled through a truncated and shifted Lennard-Jones potential:

$$U_{ij}^{LJ} = 4\varepsilon\left[\left(\frac{\sigma}{r_{ij}}\right)^{12} - \left(\frac{\sigma}{r_{ij}}\right)^{6} + 0.008742\right], r_{ij}<2.2\sigma$$

(1)

$$U_{ij}^{LJ} = 0, \qquad r_{ij}>2.2\sigma$$

The length and energy scale of this interaction sets the units. All quantities presented in the rest of this paper are expressed in terms of $\sigma$, $\varepsilon$, and $\tau = \sqrt{m\sigma^2/\varepsilon}$, where m is the monomer mass. Appropriate values for real materials are in the range of a fraction of a nanometer, a few tens of meV, and a few nanoseconds, respectively. Our model for polymers is similar to that used by Varnik et al.[10] These authors obtained the glass transition temperature for a bulk sample by fitting relaxation times obtained from diffusion data to predictions made by the mode coupling theory. They determined the bulk critical temperature of this model system to be 0.45.

Interactions between polymeric beads and surface particles are modeled through the same potential with modified length scale $\sigma_{ps}$ =0.9 and energy scale $\varepsilon_{ps}$. Two values for the strength of the attraction between chain beads and surface sites have been modeled: $\varepsilon_{ps} = \varepsilon$ and $\varepsilon_{ps} = 0.1\ \varepsilon$.

The FENE (finitely extendable nonlinear elastic) potential is used to model chain connectivity and chain grafting to the surface. Thus interactions between neighboring beads along a chain and that between grafting sites on the chains and wall atoms are modeled by means of the following potential:

$$U_{ij}^{FENE} = -\frac{1}{2}kR_0^2 \ln\left[1-\left(\frac{r_{ij}}{R_0}\right)^2\right] \qquad r_{ij}<R_0$$

(2)

$$U_{ij}^{FENE} = \infty, \qquad r_{ij}>R_0$$

Here $R_0$ equals 1.5 and the distance between the particles $i$ and $j$ in a chain at which $U_{ij}$ is at a minimum is approximately $r_{ij}=0.96$. The spring constant $k = 30$ is used to set the rigidity of a bond. It needs to be large enough to assure that bonds in the polymers do not break or cross, yet small enough to allow usage of a reasonable integration time step $\Delta t$. The diameter of the substrate particles is set to $0.8\sigma$, the nearest-neighbor distance to $0.946\sigma$. To prevent the polymers from positioning themselves along the lattice of the substrate, the distance between substrate particles differs from that between monomers along a chain

We carry out molecular dynamics simulation using a thermostat to maintain a constant temperature in the system. For beads on the polymer chains the equation of motion is

$$m\ddot{\mathbf{r}}_i = -\nabla \sum_{j \neq i} \left(U_{ij}^{LJ} + U_{ij}^{FENE}\right) - \Gamma \dot{\mathbf{r}}_i + \mathbf{W}_i(t) \quad (3)$$

Here the damping constant $\Gamma=0.4$ and $W_i(t)$ is a white noise source. The strength of the noise is related to $\Gamma$ via the fluctuation-dissipation theorem[11]. In the simulations, the equation of motion is integrated with a fifth-order Gear-predictor-corrector algorithm[13] with $\Delta t=0.005\,\tau$.

Initial states are prepared at the highest temperature. Subsequently, the film is cooled at a rate of $1/25,000\ 1/\tau$ to a new temperature setting at which it is then equilibrated for at least $40,000\ \tau$. Finally, runs are performed for up to $200,000\ \tau$. Although the data obtained from these runs fluctuate, no systematic trends due to aging were detected. Moreover cooling rates up to five times slower, did not yield detectable differences in the results.

The film has dimensions of 19 x 16 in the plane of the film. As is customary, periodic boundary conditions are used in the plane of the film ($x$ and $y$ directions). In some of the simulations, 25 percent of the chains in the film are side-grafted to the supporting surface. This yields the same ratio of thickness of the grafted layer to the radius of gyration of individual chains as reported in experimental work.[7] However, we note that we are unable to deduce the average number of grafting sites on each chain from the experimental data. In the simulations at hand, the grafted chains contain an average of 4.8 connections with the substrate. This number will be varied in future work. In order to prepare the grafted film, temporary crosslinks will be created through an algorithm described in an earlier publication[14] by one of the authors. In this algorithm, Monte Carlo moves are employed to form or break the FENE bonds between the polymer beads and the surface. The average number of connected beads per grafted chain is controlled through a parameter that models the associative attraction and controls the success rate of the Monte Carlo moves. After the system is equilibrated, the FENE bonds are made permanent.

The glass transition temperature of the thin films in this work is determined using three different approaches. The first approach consists of monitoring the film thickness as a function of temperature. Film thickness in this case is determined using the density distribution of the chain beads as a function of the distance from the supporting substrate.

At a critical temperature $T_h$ the plot of film thickness versus temperature changes slope. The method is essentially the same as that employed experimentally using ellipsometry. The second approach for determining the glass transition follows earlier work by Perera et al.[15] They obtained the specific heat-temperature curve for a binary mixture of soft disks, which is a model glass forming system. In the isothermal-isobaric ensemble at hand, the constant pressure heat capacity $C_p$ at zero pressure is given by

$$C_p = \frac{\langle \Delta E^2 \rangle}{NT^2} \quad (4)$$

where E is the total energy and N the number of beads. This approach is similar to the calorimetric method used experimentally where the DSC experiments are used to determine the glass transition by monitoring the temperature derivative of the enthalpy. For a system at equilibrium, expression (4) can be derived from the fluctuation theory of statistical mechanics. Interestingly, as Perera et al.[15] show, the correspondence expressed by this expression holds into the supercooled state, although one could argue that the system was only "locally" equilibrated in all the basins of the glassy energy landscape that were visited during the run time.

The heat capacity versus temperature data show an asymmetric peak. Two transition temperatures can be deduced from them. First, the fictive temperature $T_f$, is the temperature at which the heat capacity begins to rise. Second, the temperature denoted by $T_p$, at which the heat capacity peaks. In their simulations of liquid silica, Saika-Vovoid et al.[16] relate the observed peak in the heat capacity to changes in the potential energy hypersurface, whereas, in general, $T_f$ is related to the dynamic arrest of particles at the glass transition.

The diffusion coefficient ($D_\alpha$) of the chain beads is monitored in the third method. The characteristic time for translational $\alpha$-relaxation ($\tau_{tr}=D_\alpha^{-1}$) is then obtained for a range of temperatures above the glass transition. According to the idealized mode coupling theory (MCT), this characteristic time diverges at a critical temperature $T_c$ as $\tau \propto (T-T_c)^{-\gamma}$. Alternatively, the Vogel-Fulcher-Tammann (VFT) equation $\tau \propto e^{c/(T-T_0)}$ can also be employed to analyze the data and to obtain a VFT temperature $T_0$. In our simulations, as well as those by others[10,17,18], simulated relaxation times cover at the most one or two decades. As a result, both power-law (MCT) and exponential (VFT) fits can represent the data equally well over this limited range. For this reason, we consider an analysis based on diffusion to be less accurate than the other two approaches.

### III. RESULTS

#### A. Film thickness

For an ungrafted film with polymer-substrate interaction $\varepsilon_{ps}=1$, Fig 1a shows the bead density as a function of distance from the first layer of atoms in the substrate. The temperature equals 0.55. As has been reported before[19,20], layering is observed near the surface. It is more pronounced and extends further inward at lower temperature. At the lowest temperatures studied, layering extends throughout the film. At high temperatures (T=1.4), only one layer is observed. A bulk-like region of constant bead density occurs farther away from the surface. Ultimately, the bead density reduces to zero near the free surface of the film. The film thickness is determined as the midpoint of the region where the bead density falls from its bulk-like value to the free surface value. We find that the film thickness equals $13.3 \pm 0.2$ at $T = 0.55$. Fig. 1b shows a plot of film thickness for a range of temperatures. Linear sections of different slopes can be clearly distinguished for the liquid and glassy regions. The point of intersection of lines representing the liquid and glassy regions yields a transition temperature value $T_h$ of 0.51. We would like to point out that there is some degree of subjectivity involved in determining the value of the glass transition using this way. It depends on the temperature range chosen for representing the linear portions of the film thickness in the liquid and glassy regions.

Fig 2a compares the density profile of the ungrafted film with that of the grafted one. The curves for the two films essentially lie on top of each other. For the grafted film, the first peak in the density profile, is closer to the substrate. We attribute this to the FENE interactions between the grafted beads and atoms in the substrate. Temperature dependence of the thickness of the grafted film (shown in Fig. 2b) yields a transition temperature of 0.54, which is slightly higher than that obtained for the ungrafted film. The shift is due to a small difference in film thickness values at higher temperatures.

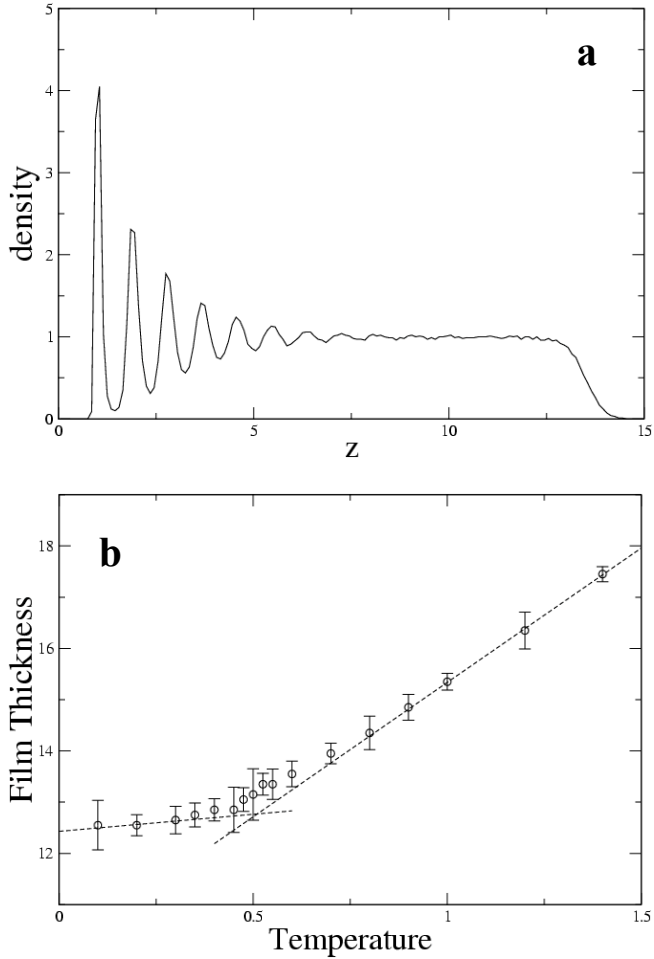

FIG 1. (a) Bead density as a function of distance (z) from the substrate for an ungrafted film with $\varepsilon_{ps} = 1$ at a reduced temperature $T = 0.55$. (b) Film thickness as a function of temperature for the same film. Fits to the data at low and high temperature are shown. Glass transition temperature is obtained from fits to the data at low and high temperature. The crossover temperature $T_h = 0.51$.

Fig. 3 displays the same data for simulations in which the LJ interaction strength between polymer and substrate is decreased to $\varepsilon_{ps}=0.1$. Data were obtained for both ungrafted and grafted films. For ungrafted films, however, all the simulated temperatures are less than 0.6, since the films separate from the substrate at higher temperatures. Density profiles at $T = 0.55$ are shown in Fig 3a. Comparison to that of the ungrafted film at $\varepsilon_{ps}=1$ reveals the expected change in layering near the substrate. For the ungrafted film, at the lower polymer-substrate attraction, layering is more or less absent, whereas for the grafted film the tendency to layer is strongly reduced. Naturally, the changes in bead packing near the surface result in differences in film thickness. Film thickness increases to 13.85 for the ungrafted film at $\varepsilon_{ps}=0.1$ and to 13.55 for the grafted film at $\varepsilon_{ps}=0.1$. Similar changes occur at other temperatures. Film thicknesses as a function of temperature at $\varepsilon_{ps}=0.1$ for both films are shown in Fig 3b. In the glassy region, data for grafted and ungrafted films are the same. By contrast, in the liquid region the ungrafted film is thicker- as was also observed in Fig 3a. The glass transition temperature for the ungrafted film, obtained from the fits shown, equals 0.49. Since we were unable to obtain data above T=0.6 for the ungrafted film, we could not obtain a fit to the film thickness data in the liquid region. Nor could we calculate the crossover temperature. At
temperatures between T=0.4 and T=0.6, the ungrafted films are thicker than the grafted films, which suggests that the glass transition temperature of the ungrafted film must be below that of the grafted film.

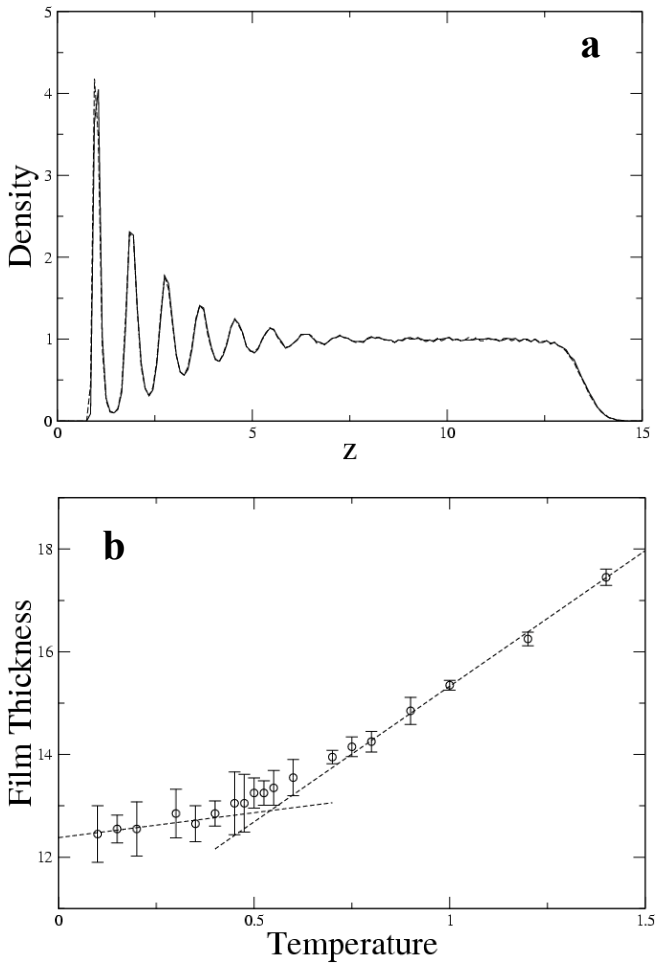

FIG 2. (a) A comparison of the bead density of the ungrafted (solid ) and grafted (long dashed) films with $\varepsilon_{ps} = 1$ as a function of the distance (z) from the substrate. (b) Film thickness as a function of temperature for grafted film, $\varepsilon_{ps} = 1$. The glass transition temperature obtained from fits equals 0.54.

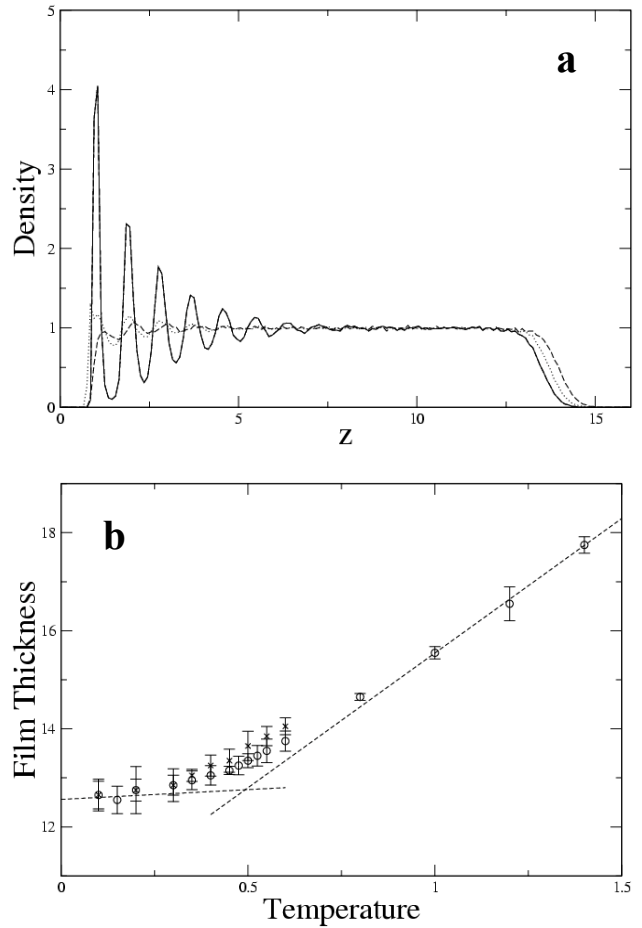

FIG. 3. (a) Bead density as a function of the distance (z) from the substrate for an ungrafted film with polymer-substrate interaction $\varepsilon_{ps} = 1.0$ (solid) and with $\varepsilon_{ps} = 0.1$ (long dashed) as well as a grafted film with $\varepsilon_{ps} = 0.1$ (dotted). (b) Film thickness as a function of temperature for an ungrafted (x) and grafted (o) film at $\varepsilon_{ps} = 0.1$. $T_h = 0.49$ for the grafted film. The ungrafted film separates from the substrates for temperatures above 0.6.

### B. Heat capacity

In Fig 4 a-d, the constant pressure heat capacity $C_p$ is plotted as a function of temperature for all four films.

Different methods have been followed in the literature to infer the glass transition temperature from calorimetric data. Glass transition temperature can be defined as the temperature at which the heat capacity as a function of temperature peaks, or as the temperature at the onset of a rise in the heat capacity.[15,16,21,22,23] We will call the temperature obtained using the latter
method the fictive temperature $T_f$ and that obtained using the former $T_p$. Data for grafted and ungrafted film at $\varepsilon_{ps}=$ 1 are shown in Figs 4a and 4b. Both types of films show an asymmetric peak in the specific heat data. The specific heat for the ungrafted film peaks at $T_p = 0.5$. For the grafted film the peak is less pronounced, but appears to be shifted slightly to the right, to $T_p = 0.55$. The temperature $T_f$ at which the heat capacity starts to rise is obtained from the crossover of fits to the data just before the onset of the rise and those during the rise. This yields $T_f = 0.32$ for the ungrafted and $T_f = 0.33$ for the grafted film. In Figs 4c and 4d we plot the specific heat versus temperature for films with the less attractive polymer-substrate interaction.

These data peak at $T_p = 0.45$ (ungrafted) and $T_p = 0.49$ (grafted). The onset of the rise of Cp is $T_f = 0.32$ (ungrafted) and $T_f$ is 0.33 (grafted). It is instructive to display the data for all four films in the same figure. From such a display, shown in Fig 5, we arrive at the following conclusions.

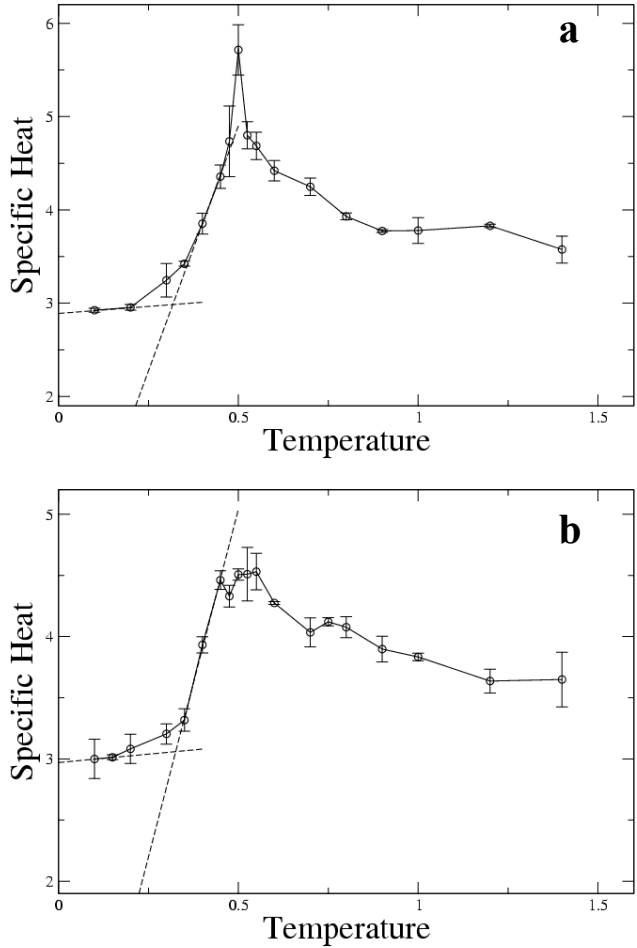

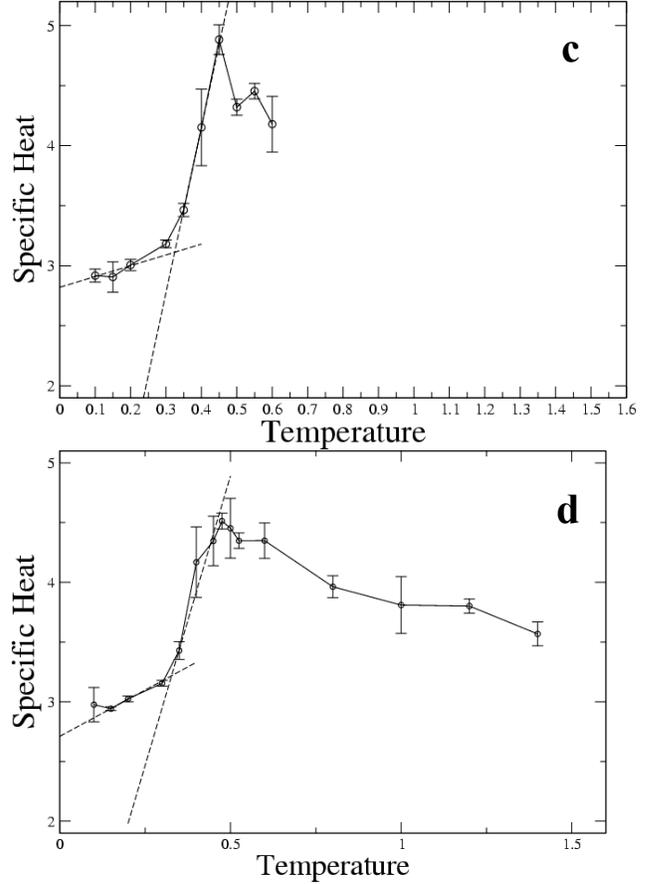

FIG. 4. Temperature dependence of constant pressure heat capacity ($C_p$) for (a) ungrafted film, $\varepsilon_{ps} = 1$ (b) grafted film, $\varepsilon_{ps} = 1$ (c), ungrafted film, $\varepsilon_{ps} = 0.1$ (d) grafted film, $\varepsilon_{ps} = 0.1$. Glass transition temperature resulting from shown fits to data equal $T_f=0.32$ (a), $T_f=0.33$ (b), $T_f=0.32$ (c), $T_f=0.33$ (d).

First, the specific heat versus temperature shows a sharp peak for ungrafted films, whereas for grafted ones the peak is much broader. Second, if the polymer-substrate attraction is lowered from $\varepsilon_{ps}=1$ to $\varepsilon_{ps}=0.1$, the position in the peak for ungrafted films shows a clear shift to lower temperatures. Third, the onset of the peak occurs at more or less the same temperature for all four films. Fourth, the slope in the data just beyond the onset of the peak appears to depend on the polymer-substrate interaction, since it is lower for higher attraction. It is insensitive to grafting, though.

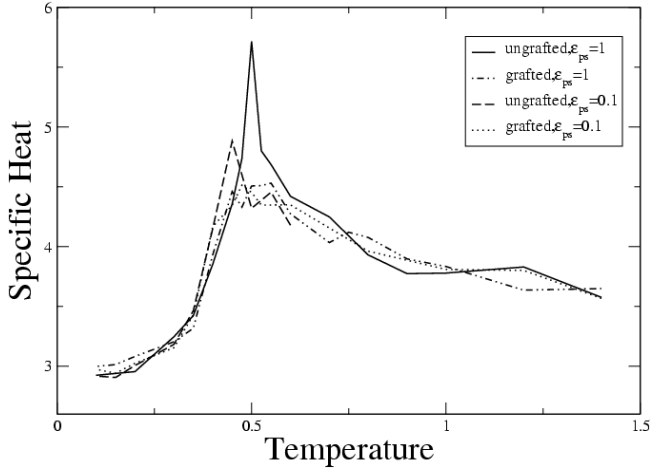

FIG. 5. Comparison of specific heat data for all four films.

**C. Diffusion**

The local translational mobility of the monomers in the films has been studied by means of the mean squared displacement (MSD) of the beads. For this purpose, the MSD is calculated by averaging the squared displacement in the *xy*-direction over all particles in the system. Displacement of the beads in the third direction (perpendicular to the plane of the film) has not been included, since the motion of the particles in this direction is influenced by the substrate and the free surface. The MSD is corrected for the center-of-mass diffusion introduced by the thermostat. For films containing grafted chains, contribution from beads belonging to the grafted chains is excluded from the MSD. Figure 6 shows the results of the MSD for different temperatures of the ungrafted film at $\varepsilon_{ps}$=1.

Several groups[10,17,18] have used MSD data to obtain the glass transition temperature from molecular dynamics simulations using mode coupling theory (MCT). Some of these simulations use models that include chemical detail[17], others, like ours, employ somewhat more coarse-grained bead-spring models of polymers[10,18] In all these studies the MSD data are employed to calculate relaxation times as a function of temperature. A critical temperature $T_c$ is defined as the temperature at which the relaxation time diverges. Although the simulations can only access a small range of relaxation times, excellent agreement with experimental results is obtained if the simulation model includes the detail necessary to make such a comparison.[17]

For the simulations at hand, the mean square displacement over time can be seen in Fig 6. A plateau regime occurs at low temperature (T=0.3), a result of the beads getting trapped in the cages formed by their neighbors. For short times, one would expect the motion to be ballistic. Although observed in our simulations, this regime is not shown. As temperature increases, the horizontal line representing a plateau regime gets shorter and is followed by a subdiffusive or $\alpha$-relaxation regime. Bead motion in this regime can be represented using a power law expression:

$$<R^2(t)> = (D_\alpha t)^\alpha \qquad (5)$$

where $D_\alpha$ is the diffusion constant. Fitting the simulation data to this expression in the time interval $t = 500$ to $t = 4000$ at different temperatures gives an average value of $\alpha = 0.46 \pm 0.04$.

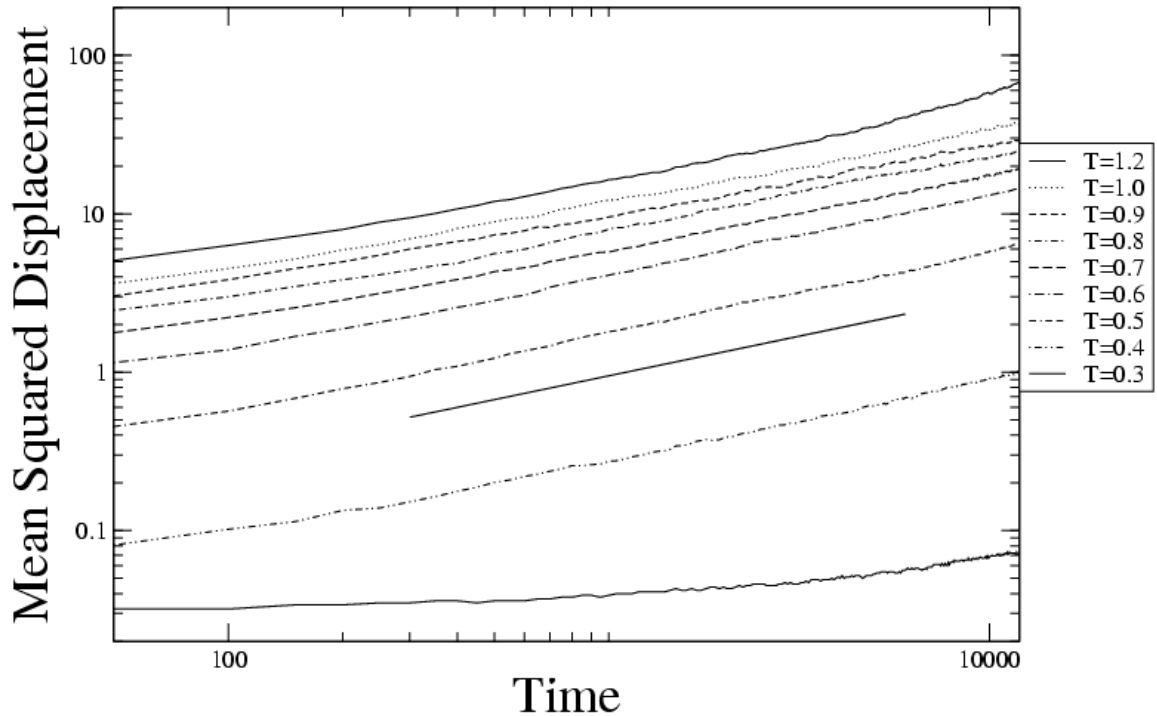

FIG. 6. Log-log plot of mean squared displacement of chain beads as a function of time in the ungrafted polymer film, $\varepsilon_{ps}$ = 1.

According to MCT, the characteristic time of the translational $\alpha$-relaxation in the sub-diffusive regime, $\tau_{tr}=D_\alpha^{-1}$, algebraically diverges at the critical temperature as[17]

$$\tau_{tr} = \frac{\tau_0}{(T-T_c)^\gamma} \qquad (6)$$

The value of $D_\alpha$ and hence the relaxation time at each temperature was determined by fitting the MSD data to equation (5) and using a value of $\alpha = 0.46$. The relaxation time $\tau_{tr}$ in Fig 7 is plotted against the temperature $T$. Fitting the data in this plot with equation (6) gives: $T_c = 0.36$ and $\gamma = 2.9$. This value of the exponent $\gamma$ compares favorably with a value of 2.1 obtained by Varnik et al.[10] for thin films and a value of 2.85 obtained by Zon and Leeuw[18] for bulk polymers. Fig 7 also show data for the grafted chains at the same polymer-substrate attraction. The results of a fit to equation (6) (not shown) are very similar: $\alpha = 0.49$, $T_c = 0.36$ and $\gamma = 3.0$.

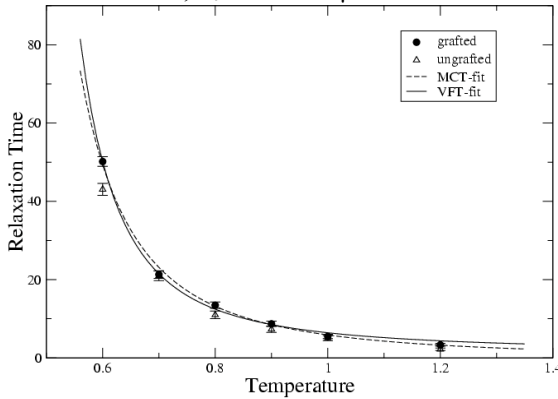

**FIG. 7.** Relaxation time as a function of temperature for ungrafted and grafted polymer films, $\varepsilon_{ps} = 1$. The long dashed line corresponds to a fit to equation (6), which is motivated by the idealized mode coupling theory. The solid line corresponds to a fit to equation (7), the VFT equation.

## IV. DISCUSSION

The glass transition is a complex, time-dependent kinetic process, hard to characterize in terms of one single transition temperature[23]. A free volume concept is often employed to describe the glass transition. As the free volume or the average unoccupied volume available for bead motion decreases, the relaxation time increases. Mode coupling theory predicts that the relaxation time diverges at the critical temperature $T_c$. Unoccupied volume by itself is not, however, an adequate measure for the state of a glassy film. Films with the same unoccupied volume, but different prepared, has been shown[24] to possess different mechanical properties. Hence, an additional mechanism must exist by which the material remembers its history. It has been hypothesized that the fluctuations in the state of local packing may play a role.[24,25] This would cause a structural

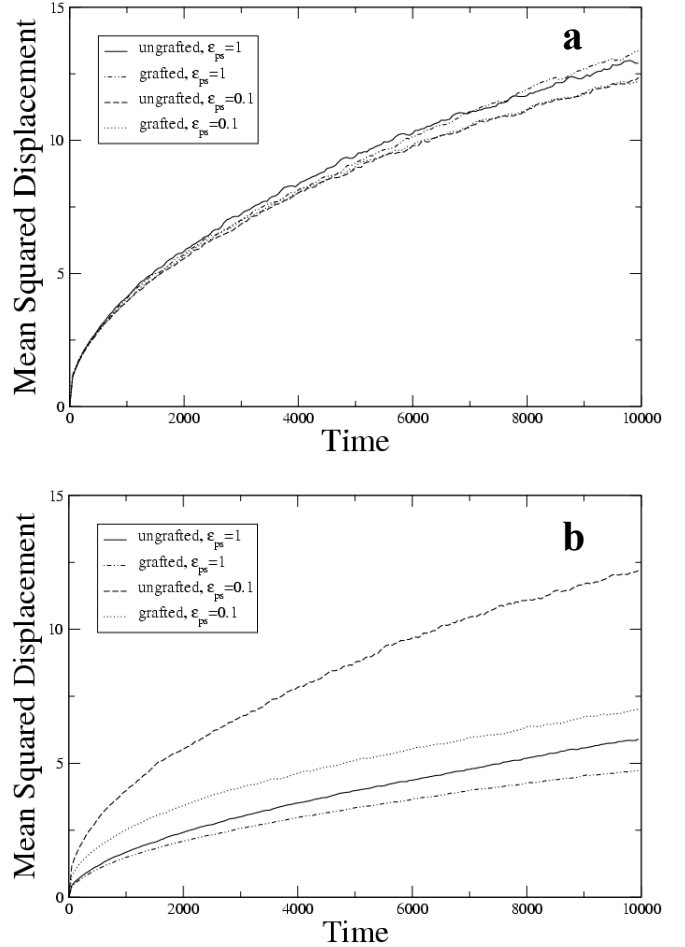

FIG. 8. Mean squared displacement (MSD) of chain beads of the polymer films as a function of time for (a) layer near the free surface (b) layer near the substrate.

|  | $T_h$ | $T_f$ | $T_p$ | $T_c$ | $T_0$ |
|---|---|---|---|---|---|
| ungrafted, $\varepsilon_{ps} = 1$ | 0.51 | 0.32 | 0.5 | 0.36 | 0.24 |
| grafted, $\varepsilon_{ps} = 1$ | 0.54 | 0.33 | 0.55 | 0.36 | 0.24 |
| ungrafted, $\varepsilon_{ps} = 0.1$ | – | 0.32 | 0.45 | – | – |
| grafted, $\varepsilon_{ps} = 0.1$ | 0.49 | 0.33 | 0.49 | – | – |

TABLE 1. Glass transition temperature values obtained using different methods. The table shows values determined from the changes in film thickness ($T_h$), as the temperature ($T_f$) at which the specific heat starts to rise, as the temperature ($T_p$) where the specific heat peaks. Critical temperatures determined from fits of relaxation times obtained from diffusion data are included as well

energy landscape with decreasing temperature has recently been investigated by Salka-Voivod et al.[16] in simulations of liquid silica. They associate it with a dynamical transition from a fragile liquid state at high temperature to a strong liquid one at low temperature. At the transition temperature $T_p$ the specific heat peaks. Hence two distinct transitions at different temperatures thickness data, match with the $T_p$ values. On the other hand, the transition temperature $T_f$, defined
by the onset of the rise in $C_p$ versus T data, is slightly lower than $T_c$, which was obtained from a fit of the relaxation time data to equation (6). Hence we hypothesize that at $T_h \approx T_p$, a transition occurs due to changes in the energy landscape, whereas the temperature $T_f$ characterizes the glass transition. Note that this implies that the change in slope in the film thickness versus temperature at $T_h$ is due to differences in local packing at high and low temperature and not due to dynamic arrest. Our data for $T_c$ (MCT) and $T_0$ (VFT) are consistent with $T_f$, especially the value of MCT critical temperature is approximately 10% higher than the glass transition temperature $T_f$.[26]

Fig 5 and Table 1 further indicate that grafting some of the chains to the substrate increases $T_p$, in agreement with experiments. Although intuitively one might have predicted the observed trend of grafting on the structural transition temperature $T_p$, the underlying mechanism is far from understood. Its explanation will most likely not be obtained without the development of a good theoretical understanding of the mechanisms responsible for the dynamical fragile-to-strong transition[16] and the accompanying specific heat anomaly.

The observed shift in the transition temperature due to grafting is quite small. The bead-spring model polymers can be mapped onto real polymers[11]. In this mapping, the value of $\varepsilon/k_B$ ranges between 300 and 500 K, depending on the experimental polymer system that is mapped. Hence, the increase in the value of the glass transition temperature as a result of chain grafting in our simulations compares to a 15-25 K increase in the value of $T_g$ for a real polymeric system. This increase in $T_g$ is much smaller than that observed in the experimental work of Tate et al.[6], who found that grafting leads to an increase in $T_g$ by as much as 55 K above it bulk value.

Since for a film thickness comparable to that used in our simulations the glass transition temperature of the film is about 30 degrees less than the bulk value, the shift in transition temperature due to grafting observed experimentally is at least 85 degrees. One possible reason for this discrepancy could be that the experimental system contained more grafted sites per chain than studied in this work (see earlier text). It is also possible that crosslinking (either temporary or permanent) within the film has played a role in further raising the transition temperature of the film in the experiments. A detailed investigation of the effects of the various grafting characteristics (number of grafted sites in the film, distribution of the grafting sites on the chains, fraction of grafted chains etc.) on the glass transition temperature will be the subject of a future publication.

Future work will also investigate molecular level structure of the films and hopefully will give an explanation for the observed dependence of $T_p$ of the films on the interaction between the polymer and the substrate, including the effects of chain grafting. Changes in molecular level dynamical[27] properties of these films across the glass transition range have recently been investigated in our laboratory as well.[28]

**Conclusions**

This paper presents a detailed investigation of the glass transition behavior of nanoscopically thin polymer films. The work is motivated by the experimental observation that grafting some of the polymer chains to the substrate can lead to a significant increase in the glass transition temperature. Such an elevation in grafting some of the polymer chains to the substrate is desirable for manufacturing processes in the microelectronics industry, such as lithography and millipede data storage technology.[29]

Although the phenomenon of glass transition has been studied for decades, developing a fundamental understanding of the molecular mechanisms underlying glass transition is still an active area of research. A strong dependence of $T_g$ on chain grafting is definitely not predicted by any of the existing theories. Simulation results such as those obtained in this work can be used to refine the theories of glass transition.

In this work, we have employed three different methods for determining the glass transition temperature. Two distinct temperatures, both of which characterize different aspects of the glass transition, can be extracted from our data. First, the structural transition temperature, either defined as the temperature $T_p$ at which the heat capacity peaks or the temperature $T_h$ at which the slope in a plot of film thickness as a function of temperature changes. In qualitative agreement with the experimental data in the literature, this temperature is found to show a dependence on both chain grafting on the substrate surface, and on the strength of the substrate-polymer interaction. Second, a critical temperature $T_f$, defined as the fictive temperature at which the specific heat as a function of temperature starts to rise. Diffusion data indicate that below this temperature, bead motion is frozen over the time scale of our computer experiments. This temperature is slightly lower than the critical temperature ($T_c$) defined by the mode coupling theory which is obtained by monitoring the temperature dependence of the relaxation time deduced from the diffusion of the beads. This method is less accurate, though, given the limited range of relaxation times that can be addressed in computer simulations. Both chain grafting and the strength of the polymer-substrate interaction are found to have no influence on the values of $T_f$ and $T_c$.

The three different methods used for determining the glass transition temperature probe different aspects of this process. A more detailed analysis that differentiates between the molecular mechanisms underlying these definitions of glass transition is currently underway in our laboratory.

## ACKNOWLEDGMENT

This research is supported by a grant from the Donors of the Petroleum Research Fund, administrated by the American Chemical Society. Moreover, Maarten v. Weert contributed to this work during a three month long student internship. He acknowledges support from the Dept. of Applied Physics of the Technical University Eindhoven, The Netherlands. We acknowledge P. F. Nealey, J. J. de Pablo, A.V. Lyulin, and M.A.J. Michels for many insightful discussions.

be distinguished. For all four types of model films studied, the values of the crossover temperature $T_h$, obtained from the film are of importance, one is due to volumetric effects, the other to structural ones.

As shown in Table 1, which summarizes our results, in our data two transition temperatures can indeed clearly

---

[*] Electronic address: abaljon@mail.sdsu.edu.
[**] Work on this project was begun and partially completed when RK held a position at Accelrys Inc. Current address of RK: Department of Chemical and Biological Engineering, U. of Wisconcin, Madison, WI.

## References and Notes

[1] Keddie J. L.; Jones R. A. L.; Cory, R. A. *Faraday Discuss*. **1994,** *98*, 219.

[2] Fryer, D.S.; Nealey P.F.; de Pablo J.J. *Macromolecules* **2000**, *33*, 6439.

[3] van Zanten, J.H.; Wallace, W.E.;Wu, W. *Phys. Rev. E* **1996,** *53*, R2053.

[4] Fryer D.S.; Peters R.D.; Kim E.J., Tomaszewski J. E.; de Pablo J. J.; Nealey P. F.; White C. C.; Wu, W. *Macromolecules* **2001**, *34*, 5627.

[5] Prucker O.; Christian S.; Bock H.; Ruhe J.; Frank C. W.; Knoll W. *Macomol. Chem. Phys*. **1998**, *199*, 1435.

[6] Tsui O. K.; Russell C.; Hawker T. P, C. J. *Macromolecules* **2001** *34*, 5535.

[7] Tate R. S.; Fryer D. S.; Pasqualini S.; Montague M. F.; de Pablo J. J.; Nealey P. F. *J. Chem. Phys.* **2001,** *115*, 9982.

[8] Yamamoto S.; Tsujii Y.; Fukuda T. *Macromolecules* **2002**, *35*, 6077.

[9] Torres J.A.; Nealey P.F.; de Pablo J. J. *Phys. Rev. Lett.* **2002**, *85*, 3221.

[10] Varnik F., Baschnagel J.,. Binder K, *Phys. Rev. E* **2002**, *65*, 21507.

[11] Kremer K.; Grest G. *J. Chem. Phys*. **1990**, *92*, 5057.

[12] M. Putz, K. Kremer, G. Grest, *Europhys. Lett*. **2000**, *49* 735.

[13] Allen M.P.; Tildesly D.J. *Computer Simulations of Liquids*; Clarendon: Oxford, 1987.

[14] Baljon, A.R.C.; Vorselaars, J.; Depuy, T.J. *Macromolecules* **2004**, *37*, 5800.

[15] Perera D. N.; Harrowell P. *Phys. Rev. E* **1999**, *59*, 5721.

[16] Saika-Voivod, I; Poole, P.H.; Sciortino F. *Nature* **2001** *412*, 514.

[17] Lyulin, A.V.; Balabaev, N.K.; Michels, M.A.J. *Macromolecules* **2002**, *35*, 9595.

[18] van Zon A.; Leeuw S. W. *Phys. Rev. E* **1999**, *60*, 6942.

[19] Khare R.; de Pablo, J.J.; Yethiraj A *Macromolecules* 1996, *29*, 7910.

[20] Bushan, B; Israelachvili J.N.; Landman U. *Nature (London)* **1995**, 374, 607.

[21] Angell C. A., Torell L. M, *J. Chem. Phys*. **1983**, *78*, 937.

[22] Efremov M. Y; Warren J. T.; Olson E. A; Zhang M.; A. Kwan T.; Allen L. H.; *Macromolecules* **2002**, *35*, 1481.

[23] Moynihan, C.T. In *Assignment of the Glass Transition*, *ASTM STP 1249*; Seyler, R.J., Ed; American Society for Testing and Materials: Philadelphia, 1994, p 32.

[24] Song, H.H.; Roe R.J. *Macromolecules* **1987,** *20*, 2723.

[25] Falk, M.L.; Langer, J.S. *Phys. Rev. E* **1998**, *57*, 7192.

[26] Colby, R.H. *Phys. Rev. E* **2000**, *61*, 1783.

[27] Benneman C.; Donati C.; Baschnagel J.; Glotzer S.C. *Nature* **1999**, 399, 24.

[28] Baljon A.R.C.; Van Weert M.H.M.; Khare R. *Phys. Rev. Lett*., in review.

[29] Vettiger P.; Cross G.; Despont M.; Drechsler .; Durig U.; Gotsmann B.; Haberle W.; Lantz M.A.; Rothuizen H.E.; Stutz R.; Binnig G.K. *Transactions on Nanotechnology* **2002**, 1, 39.